\AtBeginDocument{%
  \paperwidth=\dimexpr
    1in + \oddsidemargin
    + \textwidth
    + 1in + \oddsidemargin
  \relax
  \paperheight=\dimexpr
    1in + \topmargin
    + \headheight + \headsep
    + \textheight
    + 0.7in + \topmargin
  \topmargin = -0.5in
  \relax
  \usepackage[pass]{geometry}\relax
}

\documentclass{svjour3}                     
\smartqed 
\usepackage{graphicx}
\usepackage{amsmath}
\usepackage{amsfonts}
\usepackage{amssymb}
\usepackage{bm}
\usepackage{subfig}

\usepackage{hyperref} 

\newcommand{\beq}{\begin{equation}}
\newcommand{\eeq}{\end{equation}}
\newcommand{\beqa}{\begin{eqnarray}}
\newcommand{\eeqa}{\end{eqnarray}}

\newcommand{\chiEFT}{$\chi$EFT}
\newcommand{\chiPT}{$\chi$PT}

\begin{document}

\title{Nuclear Structure at the Crossroads\thanks{Contribution to Special Issue:
Celebrating 30 years of Steven Weinberg's papers on Nuclear Forces from Chiral
Lagrangians}}

\titlerunning{Nuclear Structure at the Crossroads}

\author{R.~J.~Furnstahl \and H.-W.~Hammer \and A.~Schwenk}

\authorrunning{Furnstahl, Hammer, and Schwenk}

\institute{R. J. Furnstahl \at
              Department of Physics, Ohio State University, Columbus, OH 43210, USA\\
              \email{furnstahl.1@osu.edu}\\
           H.-W. Hammer \at
              Technische Universit\"at Darmstadt, Department of Physics, 64289 Darmstadt, Germany\\
              ExtreMe Matter Institute EMMI and Helmholtz Forschungsakademie Hessen f\"ur FAIR (HFHF), GSI Helmholtzzentrum f\"ur Schwerionenforschung GmbH, 64291 Darmstadt, Germany\\
              \email{Hans-Werner.Hammer@physik.tu-darmstadt.de}\\
           A. Schwenk \at
              Technische Universit\"at Darmstadt, Department of Physics, 64289 Darmstadt, Germany\\
              ExtreMe Matter Institute EMMI, GSI Helmholtzzentrum f\"ur Schwerionenforschung GmbH, 64291 Darmstadt, Germany\\
              Max-Planck-Institut f\"ur Kernphysik, Saupfercheckweg 1, 69117 Heidelberg, Germany\\
              \email{schwenk@physik.tu-darmstadt.de}}

\date{Received: date / Accepted: date}

\maketitle

\begin{abstract}
Steven Weinberg's seminal papers from 1990--92 initiated the use of effective field theories (EFTs) for nuclei.
We summarize progress, priorities, and open questions for nuclear EFT developments based on the 2019 INT program ``Nuclear Structure at the Crossroads.''
\end{abstract}

\section{Overview}
  \label{sec:overview}
  
An effective field theory (EFT) exploits the hierarchy of scales in a physical system to  provide a powerful framework for controlled expansions of experimental observables. 
Steven Weinberg, more than anyone else, was the creator of EFT in its general form~\cite{Weinberg:1978kz,Weinberg:2009bg}.
He sowed the seeds for diverse implementations of EFT~\cite{HARTMANN2001267,Rivat:2020amd}, and he was the instigator of first chiral perturbation theory (\chiPT) and later chiral EFT (\chiEFT) for nucleon interactions.
The advent of \chiEFT\ transformed nuclear structure theory. 

The development of $\chi$PT in the early 1980's~\cite{Gasser:1983yg,Gasser:1987rb} did not significantly impact the theoretical treatment of nuclei (that is, nuclear structure and reactions) for the next decade because there was no extension to more than one nucleon, for which perturbation theory is inadequate (e.g., there are bound states).  Weinberg invented the way forward in 1990--1992 in his seminal papers ``Nuclear forces from chiral Lagrangians''~\cite{Weinberg:1990rz}, ``Effective chiral Lagrangians for nucleon--pion interactions and nuclear forces''~\cite{Weinberg:1991um}, and ``Three-body interactions among nucleons and pions''~\cite{Weinberg:1992yk}.  He showed how to systematically incorporate contact interactions between nucleons and higher-order pion exchanges into a chiral Lagrangian such that the derived potential could be used in the many-nucleon Schr\"odinger equation to describe nucleon-nucleon scattering and finite nuclei.  Further, he provided a theoretical explanation for the phenomenologically observed hierarchy of many-body forces and a systematic way to calculate three- and higher-body forces.  
These papers have had an enormous impact.

Three decades after Weinberg's groundbreaking papers on \chiEFT, the five-week Institute for Nuclear Theory (INT) program in 2019 entitled ``Nuclear Structure at the Crossroads'' was a meeting place for different approaches to nuclear structure and reactions. 
In dictionaries one can find various definitions of crossroads: 
 \begin{quote}
   ``A point at which a crucial decision must be made that will have far-reaching consequence.''
 \end{quote}
 \begin{quote}
   ``A situation that requires some important choice to be made.''
 \end{quote}
The operational definition for the INT Crossroads program was consistent with these:  a need to identify the priorities in moving forward in low-energy nuclear theory, particularly with respect to the role of EFT. 
The program focused on key questions addressing open issues and priorities for future research.
The participants are listed in Sec.~\ref{sec:participants} and the talks are available at the INT website \url{https://www.int.washington.edu/PROGRAMS/19-2a/}.   

Workshop participants confronted an extended list of intertwined issues, each with multiple sub-issues:
       \begin{itemize}
       \item Degrees of freedom (DOFs) [pionless or pionful, including $\Delta$ isobars or not, clusters, collective DOFs];
       \item Hamiltonians/currents for ab initio calculations [power counting, model independence, expectations for leading order (LO), renormalization group (RG) invariance, regulator artifacts]; 
       \item Uncertainty quantification (UQ) [EFT truncation errors, Bayesian parameter estimation, model checking, model selection and evidence]; 
       \item Ab initio methods and tools [future of many-body methods, eigenvector continuation, emerging technologies];
       \item Lattice QCD [implementation for more than one nucleon, implications for nuclear structure and reactions];
       \item Energy density functionals (EDFs) as EFTs [new EDF constructions, EDF vs.\ ab initio];
       \item Reactions and structure [consistency; improving phenomenological calculations];
       \item Experimental impact also beyond nuclear physics [neutrino experiments, dark matter direct detection; multimessenger astronomy].
       \end{itemize}
In the present contribution, we have summarized the ideas for proceeding that emerged from presentations and discussions at the INT Crossroads, along with selective updates. 
This is not intended to be an exhaustive treatment but a summary of the main points from the INT program; for more extensive background, references, and updates, we refer the reader to recent reviews~\cite{Machleidt:2016rvv,Epelbaum:2019kcf,Hammer:2019poc,Hergert:2020bxy,Hebeler:2020ocj,Tews:2020hgp} and to the 2021 INT program ``Nuclear Forces for Precision Nuclear Physics'' \url{https://sites.google.com/uw.edu/int/programs/21-1b}.

\section{What counts as a nuclear effective field theory?}
  \label{sec:EFT}
  
\textbf{Overview.}
While there is consensus on what constitutes a perturbative EFT \cite{Weinberg:1978kz,Gasser:1983yg,Georgi:1994qn,Kaplan:1995uv}, the debate on nonperturbative EFTs  is ongoing. 
INT Crossroads talks by van Kolck, Ekstr\"om, Hebeler, K\"onig, Krebs, Long, Papenbrock, and S\'anchez S\'anchez led to vigorous discussions on the requirements from the perspective of nuclear EFTs; we summarize here the key points.

Perturbative EFTs require the most general effective Lagrangian (i.e., a complete operator basis) consistent with all relevant symmetries and a power counting scheme that classifies the terms in the effective Lagrangian according to their importance at low energies. 
This ensures a model-independent description.
In such EFTs the number of diagrams at each order in the EFT expansion is finite. 
As a consequence, observables and amplitudes can be calculated in strict perturbation theory without any non-trivial resummation of diagrams. 
After renormalization, the observables are independent of the regulator, up to higher-order corrections for a wide range of cutoffs (RG invariance).  
For the EFT to be useful in practice, a semi-quantitative result at leading order is required or else it is not possible to converge to the correct result in a perturbative expansion.

In nonperturbative EFTs, in contrast,  the number of diagrams that contribute at leading order is infinite, requiring a resummation  of these diagrams. 
The same situation may hold in higher orders as well, although it does not have to. 
The wish list for features of a nonperturbative EFT includes all the properties of perturbative EFT: a complete operator basis (for model independence), a consistent power counting with successively decreasing contributions at higher order, and a semi-quantitative result at leading order. 
However, this list may be guided too much by the expectations from the perturbative case. 
The resummation of diagrams often introduces higher-order divergences that cannot be renormalized at the given order. This limits the range of cutoff variations severely. It is not obvious whether all requirements from the wish list can be satisfied and there is not consensus on whether all are truly necessary for an EFT. 

\textbf{Progress and priorities.}
The discussions at the program showed that the requirement of a complete operator basis appears to be universally accepted. 
This conclusion stems from Weinberg's folk theorem: ``If one writes down the most general
possible Lagrangian, including all terms consistent with assumed symmetry
principles, and then calculates matrix elements with this Lagrangian to any
given order of perturbation theory, the result will simply be the most general
possible S-matrix consistent with analyticity, perturbative unitarity, cluster
decomposition and the assumed symmetry principles.''~\cite{Weinberg:1978kz}. 
A corollary of this folk theorem is that regulators ideally should not break any of these symmetries; if it is nevertheless broken, the symmetry must be restored at the accuracy of the EFT expansion, e.g., by adding appropriate counterterms. 
This is particularly relevant for the consistency of interactions and currents in \chiEFT\ \cite{Filin:2019eoe}.  
Moreover, it should be emphasized that an effective Lagrangian with redundant operators, i.e., an overcomplete operator basis, may make uncertainty quantification (UQ) problematic and such terms should generally be eliminated.

The requirement of a working power counting appears to be universally accepted as well. The power counting provides the basis to make EFTs useful by specifying which terms have to be included for a given accuracy.  
In doing so it establishes an \textit{a priori} truncation error based on the omitted terms. 
There are various ways at different levels of sophistication to deal with EFT errors quantitatively. 
Bayesian methods have been introduced recently to low-energy nuclear physics~\cite{Furnstahl:2015rha,Melendez:2017phj} and their application to EFT UQ were widely discussed at the program (see next section).

In perturbative EFTs the cutoff can often be taken arbitrarily large, but in practice it is sufficient to take it to the scale of new physics. Cutoff effects are then of the same order as higher-order terms. However, it is not always advisable to do this, because the cutoff dependence can also be exploited to extract physics. 
There are some examples of nonperturbative EFTs with contact interactions (where cutoff dependence was used to exhibit discrete scale invariance and the Efimov effect) as well as the KSW scheme for perturbative pions where this was done~\cite{Hammer:2019poc}. 
In these theories, only the leading order is nonperturbative while all higher-order interaction are included in perturbation theory.
        
In the Weinberg scheme the power counting is for the effective potential, which is not observable. 
This potential is then iterated to all orders in the solution of the nuclear few- or many-body problem. 
As a result the cutoff can only be varied over a relatively small range. 
The reason for this limitation is well understood as discussed above and only its interpretation is under debate. 
However, recent Bayesian analyses suggest that the expansion of actual observables may be fully consistent with the power counting. 
These analyses are also able to detect inconsistencies in a given EFT. 
Thus (at least) in practice, Weinberg counting may lead to a consistent EFT that allows for systematically improvable predictions.
A priority is to test whether different scales and schemes predict consistent observables within the theoretical uncertainties.

\textbf{Open questions.}
While there is consensus on the requirement of a complete operator basis and a working power counting scheme, there are open questions regarding the other possible requirements \cite{Hammer:2019poc}. 
To resolve the debate, all the different approaches should be applied to one or more benchmark problems~\cite{SanchezSanchez:2017tws,Wu:2018lai,Epelbaum:2018ogq,Yang:2020pgi}. 
In each case an analysis of several orders in the expansion of observables, ideally using statistical methods, should be carried out for robust conclusions.

How important is a semi-quantitative result at leading order~\cite{Luna:2019ufu,Schmickler:2019ewl}? It is certainly important for schemes that only treat the leading order nonperturbatively and treat all higher orders in perturbation theory. This is similar to the question of expanding around the correct ground state in many-body systems. If one starts with the wrong qualitative behavior, this cannot be corrected in perturbation theory. 
So this issue is to be connected to the question of a consistent power counting. A quantitative assessment can again be done by studying different orders in the expansion for their consistency, e.g., using Bayesian methods.

\section{Using EFT to quantify uncertainties in low-energy nuclear physics}
  \label{sec:UQ}

\textbf{Overview.}
There is consensus among scientists studying low-energy nuclear physics that the consistent quantification of all uncertainties, from both experiment \emph{and} theory, is essential for progress in the precision era~\cite{ISNET}.
The theoretical uncertainties include errors from the calculational methods, from estimating parameters in the model, and from incomplete or imperfect models. The latter is known as model discrepancy. 
The controlled power counting of EFTs enables estimates of the EFT model discrepancy, which in particular means the truncation error from working to a finite order in the EFT expansion.
Bayesian statistical methods are well suited for combining EFT truncation errors with other uncertainties from data (both stochastic and systematic), methods, and parameter estimation.
Here we highlight progress, priorities, and open questions stemming from presentations at the INT Crossroads program by Drischler, Ekstr\"om, Hebeler, K\"onig, Roth, Rupak, and Wesolowski,  and subsequent discussions with other participants.

\textbf{Progress and priorities.}
Low-energy nuclear physics has gone rapidly in the last few years from limited consideration of theory errors to increasingly sophisticated uncertainty quantification.   
Epelbaum, Krebs, and Mei{\ss}ner (EKM)~\cite{Epelbaum:2014efa,Epelbaum:2014sza} suggested for \chiEFT\ a protocol to use the order-by-order convergence pattern to estimate the EFT truncation error.
This is an alternative to comparing observables calculated with different regulator parameters, which has been found to be problematic for estimating uncertainties in \chiEFT\ with Weinberg counting.
The EKM approach has been formalized and generalized within a Bayesian framework~\cite{Furnstahl:2015rha,Melendez:2017phj} with a steady stream of subsequent refinements and applications.

Along with this progress there are clear priorities.
Full Bayesian parameter estimation should be applied to NN+3N \emph{and} $\pi$N interactions in \chiEFT, including the EFT truncation errors, method errors, and uncertainties from the experimental measurements used in the fit.
It is important to establish correlation matrices between the sectors and to test that there is not overfitting.
The uncertainties from parameter estimation should be consistently propagated to nuclear observables such as energy spectra and nuclear radii, fully accounting for correlations, and combined 
consistently with the theory errors in calculating those observables. 
A Bayesian framework makes it possible to learn posteriors for the EFT expansion parameters and related quantities (e.g., correlation lengths in energy or density).
This statistical analysis should be applied to different observables to test for consistency and look for anomalies.
Bayesian model checking for the convergence pattern should be applied to the full range of \chiEFT\   interactions  (Weinberg counting)  and for nuclei and nuclear matter.
To help resolve the ongoing debate on \chiEFT\ power counting schemes, Bayesian UQ should be applied to RG-invariant formulations for comparison to Weinberg counting.
Moreover, eigenvector continuation can be used for constructing an efficient and accurate emulator for nuclear observables, thereby enabling uncertainty quantification in multi-nucleon systems~\cite{Konig:2019adq}.

Pionless and cluster EFT provide valuable test cases of RG-invariant EFT for nuclear properties.
A priority is to apply the full range of Bayesian UQ for observables calculated in pionless EFT.
This includes testing the estimate of truncation errors from convergence patterns and comparing to estimates from RG dependence. 
Ultimately one would like to apply Bayesian model mixing to pionless and \chiEFT\ to take advantage of their overlapping domains of validity 
       
Some aspects of Bayesian statistical analysis that are well developed in other contexts, such as model selection~\cite{Premarathna:2019tup}, model mixing~\cite{Phillips:2020dmw}, and experimental design~\cite{Melendez:2020ikd} are still frontiers for nuclear EFT.
Early results exist but there is much to do.
Another new tool under development is the use of (global) sensitivity analysis~\cite{Ekstrom:2019lss} to quantify the impact of different operators on observables. 

\textbf{Open questions.}
While progress in UQ for EFTs, and for \chiEFT\ in particular, has been substantial, there are yet-to-be-resolved questions about developing and applying statistical methods to optimize predictions and test consistency.
Here are some of them.

How to best choose the experimental data to be input for the estimation of LECs is not clear. 
In principle it should not matter but in practice using only few-body data typically leads to deficiencies in binding energies and/or radii of larger nuclei. 
These are remedied by including information from larger $A$ nuclei in the fit, but this may in turn degrade the predictions for few-body scattering.
Is it legitimate to directly fine-tune selected observables (e.g., binding energies and radii) to achieve good phenomenological interactions or must this only be done within EFT uncertainties to maintain desirable EFT features?

A related question is how to accommodate fine-tuning that might be needed to have useful predictions.
For example, in calculating a nuclear reaction the threshold energy may need to be very precise to have any hope of a good description.
Could this be done through renormalization conditions with a subsequent adjustment of the power counting?

The multiplicity of different EFT formulations for nuclear phenomena invite the question of whether statistical methods of model selection can be used to decide which is ``best'' or if that is even a well-posed question.
A particular issue is the inclusion of $\Delta$'s in \chiEFT; there are phenomenological successes with and without.
Further, it is as yet unknown whether \chiEFT\ with conventional Weinberg power counting implemented with different regulators are statistically consistent.
Should one use Bayesian evidence to compare or formulate as mixture models?
The optimal approach might be a form of Bayesian model mixing, because the different EFT formulations should be valid within a defined domain, so the question is how to leverage them such that one enhances the overall predictive power rather than merely pitting one against another.

\section{New frontiers in the developments of nuclear forces}
  \label{sec:frontiers}

\textbf{Overview.} 
While Weinberg's papers laid out a path to the systematic development of nuclear forces, in practice there are complications, such as the need for regulators, which have led to as-yet unresolved issues. 
These are particularly evident as it has become possible to test interactions in larger nuclei.

In parallel to the description of nuclear structure, a major goal is the precision calculations of electroweak properties and reactions of nuclei with well-defined error estimates.
Apart from nuclear physics itself such predictions are relevant for the description of astrophysical processes and searches for physics beyond the standard model.
The consistency of nuclear forces and the currents for external probes in nuclear EFT is an important technical challenge. For example, if the regulators used for currents and interactions are not consistent, this can lead to artifacts. In the worst case, the power counting can be violated, leading to an irregular convergence pattern and an underestimation of the EFT uncertainties. 
Talks on the status and prospects for EFT interactions and currents given at the INT Crossroads program by Acharya, Bacca, Coello Perez, Hebeler, Hoferichter, Holt, Illa, Kievsky, Krebs, Lovato, Men\'endez, Pastore, Platter, Rupak, Roth, Schmickler, van Kolck, and Yao stimulated illuminating discussions.

An ultimate goal in the nuclear community is a determination of  the LECs in EFT from Lattice QCD. This would enable truly ab initio predictions of nuclear physics observables beyond the lightest systems directly from QCD. Apart from computational challenges, the main obstacles are the lack of lattice simulations near the physical pion mass and an incomplete understanding of systematic uncertainties. 
This is exemplified by qualitative discrepancies between different lattice approaches.
Talks on lattice QCD and nuclear forces  were given by
Aoki, Bedaque, Hiyama, Illa, Nicholson, Parre{\~n}o, Shanahan, Wagman, and Zhang, leading to extended discussions among all participants about the way forward.

\textbf{Progress and priorities.}
Progress on nuclear Hamiltonians has included a controlled treatment of the important inputs from $\pi N$ scattering~\cite{Hoferichter:2015hva} and the development of new chiral interactions with $\Delta$'s.
But at the same time, the situation with regulators is unclear for calculations of nuclear energies and radii, despite efforts to minimize regular artifacts by using specific functional forms~\cite{Epelbaum:2019kcf}.
Calculations with accurate two-body potentials with non-local regulators comparing three-body regulators, non-local (NL) versus local (L), with the same 3-body parameter-fit protocol 
show NL-NL can describe energies and radii for medium mass nuclei while NL-L fails.
For purely local potentials (i.e., L-L) without $\Delta$'s, 
artifacts arise from Fierz ambiguities (in particular the choice for the short-range operators in 3N forces).
This is mitigated for harder interactions (larger cutoffs), for which good energies and radii up to oxygen have been demonstrated.
In all cases, pushing to larger $A$ as well as extending to higher order in the EFT expansion can help clarify the situation.

Effective field theories provide a well-defined scheme to obtain consistent coupling to gauge or other external fields. In general, there are interactions generated by gauging the theory, i.e., replacing ordinary derivatives by covariant derivatives as well as local gauge-invariant interaction terms. The latter contributions are typically missing in model calculations but emerge naturally in an EFT based on the most general effective Lagrangian.
Significant progress has been made on consistently regulated currents~\cite{Krebs:2020pii} (that is, consistent with chiral symmetry), although these are not yet universally adopted.
The critical role of two-body currents has been shown for precision calculations of 
electromagnetic transitions and magnetic moments~\cite{Bacca:2014tla} and for neutrino reactions~\cite{Lovato:2017cux}.
The continued development of technology to apply these currents and testing them widely is a priority.

The reach of the unitary limit in nuclei is an open question. While its role in light nuclei up to $A=4$ is well established and naturally explains universal correlations such as the Phillips and Tjon lines~\cite{Hammer:2019poc}, the relevance in heavier systems is under debate~\cite{Gattobigio:2019omi,vanKolck:2020llt}. Pionless EFT, which is organized as an expansion around the unitary limit, seems to fail binding $^{16}$O at leading order~\cite{Contessi:2017rww,Yang:2020pgi}.
Whether the four-body force at NLO~\cite{Bazak:2018qnu} can correct this remains to be seen.
Further progress has been made in understanding the scale introduced by long-range Coulomb interactions, which leads to an enhancement of effective range effects~\cite{Luna:2019ufu,Schmickler:2019ewl}.

A complementary effort to the application of chiral forces and currents is being made in lattice QCD.
An area where lattice QCD can have a particularly big impact is the determination of LECs that are difficult to obtain from experiment because data is scarce or non-existent. 
This applies, for example, to the strangeness sector and hypernuclei~\cite{Illa:2020nsi}. 
If the NN$\Lambda$ force could be calculated, this would help to understand what happens in neutron stars at higher densities. Moreover, hyperon-nucleon interactions from lattice QCD can also be used in few-body calculations of hypernuclei~\cite{Hiyama:2019kpw}.
Another opportunity is LECs for three-body forces in the $T=3/2$ sector and three-neutron observables.
In practice, however, the goal of LECs from lattice QCD has only been reached for a few specific counterterms that are relatively insensitive to the pion mass. 
Electroweak physics is an area where lattice QCD can make a difference already by determining unknown matrix elements~\cite{Shanahan:2017bgi} and gauge-invariant counterterms whose coefficients are not constrained by symmetry. Another example is the physics of the EMC effect, where calculations of polarized processes can distinguish between model calculations~\cite{Winter:2017bfs}.

There are two main approaches to lattice QCD for multi-baryon systems and these have shown persistent disagreements~\cite{Tews:2020hgp}.
The HALQCD approach derives a nonrelativistic potential in a particular scheme and uses it to calculate observables~\cite{Aoki:2017byw}.
The L\"uscher method extracts observable quantities from the finite-volume dependence of energy levels on the lattice.
This method emphasizes the importance of applying lattice QCD for hadrons to identify combinations of interpolating fields that broaden the window in Euclidean time where one is dominated by the desired state.
A high priority is to resolve the discrepancies between the lattice QCD results of multi-baryon observables.
Possible flaws have been identified for both the HALQCD and L\"uscher-formula approaches.  
It is clear that a full accounting of systematic errors is still needed to resolve these issues; the planned application of both methods to the same lattice configurations will be an important step.
  
\textbf{Open questions.}
We have highlighted frontier research on nuclear EFT interactions and currents, where there is both progress and puzzles. Some of the open questions:
Are there physical motivations for picking clever regulators still to be explored?
Is there an alternative power counting not yet considered?
Are $\Delta$'s necessary for a systematic description of nuclei?
How far is the reach of the unitary limit in nuclei?
What are the high priorities for the matching of lattice data to EFT-based calculations? Which matrix elements does the ab initio community need to gear up to calculate?

\section{Advances in many-body theory for nuclei and nuclear matter}
\label{sec:advances}

\textbf{Overview}
Nuclear theory has made impressive advances in the development of ab initio methods applicable from light to medium-mass nuclei up to $A \sim 100$ and to nuclear matter~\cite{Hergert:2020bxy,Hebeler:2020ocj,Holt:2019gmc}. Moreover, EDFs have been developed with a recent focus on theoretical uncertainties and EFT ideas for the construction of EDFs.
Talks on this topic were given by Afanasjev, Bacca, Barbieri, Barnea, Bertsch, Bogner, Boulet, Engel, Ekstr\"om, Han, Hebeler, Hergert, Holt, Horowitz, Lacroix, Lonardoni, Lovato, Men\'endez, Papenbrock, Pastore, Piekarewicz, Robin, Roth, Stroberg, Tews, Wellenhofer, Wirth, Yao, and Zhang, and were supplemented by intense follow-up discussions.

The central challenges for structure include the development of accurate nuclear forces for many-body applications, extending methods to incorporate more correlations and enable more accurate treatments of open-shell and weakly bound nuclei, as well as a full understanding of theoretical uncertainties, ranging from the propagation of the input uncertainties, to EFT truncation errors, and many-body method errors.
Calculations have largely focused on using chiral NN and 3N interactions up to N$^2$LO or N$^3$LO based on Weinberg power counting (see Refs.~\cite{Hoppe:2019uyw,Huther:2019ont} for recent N$^3$LO NN+3N calculations), with initial work exploring $\Delta$-full interactions~\cite{Jiang:2020the}. For lighter nuclei, calculations also exist based on pionless EFT (with connections to lattice QCD simulations), and there are pioneering calculations using halo EFT and EFT for heavy nuclei \cite{Hammer:2017tjm,CoelloPerez:2015ksi}. Exploring pionless EFT for larger $A$ and nuclear matter also explores the question of how close nuclei and nuclear matter are to the unitary limit \cite{Kievsky:2018xsl}.

To go beyond nuclear structure observables, consistency between
structure and reaction calculations is universally regarded as essential. 
The main ab initio progress for nuclear reactions has been in few-body/light systems. 
Calculations of electroweak properties (magnetic moments, electromagnetic transitions, beta decays) have shown that two-body currents play a key role~\cite{Bacca:2014tla,Gysbers:2019uyb} and that there is a tradeoff between the contributions of such currents and correlations in the many-body wave functions.
This tradeoff can be exploited using the renormalization group to take full advantage of factorization from scale separation. Exploring this further in short-range correlation experiments is an important future topic~\cite{Cruz-Torres:2019fum,Tropiano:2021qgf}.

New developments for nuclear matter, including using many-body perturbation theory (MBPT) and quantum Monte Carlo (QMC), have been based on chiral NN and 3N interactions up to N$^2$LO or N$^3$LO~\cite{Drischler:2017wtt,Lonardoni:2019ypg}, including Bayesian uncertainty estimates~\cite{Drischler:2020hwi}.
Saturation and symmetry energy properties emerge naturally from these calculations but the connections between nuclear matter saturation and ab initio predictions of nuclei have not been solidified.

Nuclear EDFs have focused on uncertainty quantification and extensions to fission, neutron stars, and other applications.
Less progress has been made on new types of functionals beyond derivations using the density-matrix expansion.

\textbf{Progress and priorities.}
New ideas and technical developments have helped push the precision and reach of ab initio methods to new domains (see also Sec.~\ref{sec:opportunities}).
While once it seemed that computational costs would limit ab initio calculations to the low end of the nuclear chart,  
the feasibility of quantitative calculations of nuclei as large as neutron-rich tins has been demonstrated~\cite{Miyagi:2021pdc}.
Priorities include the development of tractable methods to enable higher-body-operator calculations (e.g., the inclusion of IMSRG(3) contributions~\cite{Heinz:2021xir} or four-body interactions); and the extension of emulators for nuclei to enable full UQ. A continued focus will also lie on key nuclear matrix elements for neutrino physics~\cite{Engel:2016xgb}.

The advances in calculational capabilities are providing impetus to better exploit EFT methods and implementations.
Priorities include exploring alternatives to Weinberg power counting for \chiEFT\ and pushing pionless EFT to higher $A$, which can help establish universal features in nuclei.
Similarity renormalization group (SRG) methods have played an important role softening \chiEFT\ Hamiltonians~\cite{Bogner:2009bt,Furnstahl:2013oba}, but the SRG has yet to be applied to pionless EFT.
Also on the wishlist is to explore more matching calculations between different EFTs, e.g., ab initio based on \chiEFT\ to halo EFT or EFTs with collective DOFs. Matching EFTs at different scales can lead to a tower of EFTs that covers all aspects of nuclear physics.
Even without a rigorous EFT, developing EFT-motivated ideas can systematically improve  phenomenological models.
A particular opportunity is the use of EFT to improve nuclear EDFs.
A DME implementation of long-range chiral interactions has shown improvement over pure Skyrme functionals, but there are many open questions to address~\cite{NavarroPerez:2018tme,Zurek:2020zys}. 
An alternative program for DFT from EFT is to apply field-theoretic background-field methods for dealing with symmetry breaking and restoration~\cite{Furnstahl:2019lue}.  
Making these connections can help to better understand the relationship between nuclear matter and finite nuclei.

Nuclear matter theory is not only providing insights to nuclear forces and constraints to new EDFs, but is playing a key role for multimessenger observations of neutron stars (see, e.g., Ref.~\cite{Capano:2019eae}). Improving the accuracy at saturation density and higher, and extending calculations to finite temperature~\cite{Keller:2020qhx} and proton fraction will provide pivotal constraints for the properties of neutron stars, as almost all equation of state models at low densities are rooted in many-body calculations based on \chiEFT\ interactions. 

\textbf{Open questions.}
Along with the successful extension of many-body theory for nuclei and nuclear matter come new questions.
What are the limits in mass number and density for applying pionless and \chiEFT?
Where are four-body forces important, if anywhere?
Can EFTs with factorization methods describe short-distance/high-momentum probes?
How can EFTs be used to explore the emergence of low-resolution or collective properties from more complex calculations at higher resolution?
How can we interface ab initio calculations with EDFs?
Will mass table calculations be via ab initio methods in the future?
Are there new ideas beyond the density-matrix expansion for building ab initio EDFs based on EFTs?
What is the best way to include information from multimessenger observations for low-energy nuclear physics?
Are there key many-body systems or benchmarks like the unitary Fermi gas?

\section{New opportunities from emerging technologies}
\label{sec:opportunities}
  
\textbf{Overview.}
In the three decades since Weinberg launched nuclear EFT there has been an exponential increase in computing power along with dramatic improvements in algorithms for computational few- and many-body applications.
These advances have greatly pushed the precision and reach of low-energy nuclear theory.
Looking to the future, there are potential new opportunities for nuclear structure from emerging technologies.
These include machine learning~\cite{Raghavan:2020bze}, quantum computing~\cite{Dumitrescu:2018njn,Lu:2018pjk,Klco:2020rga,Roggero:2020sgd}, computing technologies such as tensor networks~\cite{Tichai:2020jpk,Tichai:2021rtv,Zhu:2021pis}, and  automatic code generation~\cite{Arthuis:2018yoo,Arthuis:2020tjz,Tichai:2021ewr}.
Talks at the INT Crossroads program that specifically addressed such opportunities were given by Bedaque, Duguet, Drischler, Hergert, Lee, and Lovato, and there were many additional contributions during the discussion times.

\textbf{Progress and priorities.}
Frontier computing technologies are starting to make an impact on nuclear structure and reaction calculations.
Large-scale sampling techniques can be exploited with automatic differentiation.
Automatic code generation are addressing problems such as the need to identify Feynman diagrams and convert them to numerical code  and handling the combinatorics for many-body methods. 
Tensor network methods can tame the complexity and memory requirements of many-body methods such as coupled cluster and in-medium SRG, help deal with the sign problem, and provide new perspectives on entanglement. 

Emulators are computer models (or surrogates) for many-body calculations that are too expensive to run repeatedly, as needed for full UQ.
Gaussian processes provide versatile emulators with built-in error estimates.
A new technology called eigenvector continuation (EC) offers fast and accurate emulation by generating a spectacularly effective basis for variational formulations, including both bound state and scattering~\cite{Frame:2017fah,Demol:2019yjt,Furnstahl:2020abp}.
EC can enable full statistical analyses for nuclear calculations that were previously computationally prohibitive~\cite{Konig:2019adq,Ekstrom:2019lss} and also provide improved many-body expansions.

The methods from quantum information science offer the potential to provide new insights to nuclear structure, and correlated many-body states in particular, as well as solving effectively intractable problems.
This could include data mining of configurations for nuclear many-body systems, in analogy to frontier efforts for spin systems in condensed matter physics.
Machine learning is being applied for the ill-posed problem of inverting Laplace transform to extract the real-time response of electroweak scattering from QMC calculations in Euclidean time~\cite{Raghavan:2020bze} and to find improved variational wave functions. 
The sign problem in QMC is being handled by identifying complex manifolds for path integration.

While not directly addressed in the scope of the INT Crossroads program, quantum computing was a regular part of discussions.
If the potential for a qualitative enhancement in computational capabilities via quantum computing is realized, new paradigms for nuclear structure calculations could emerge. 
While the technical hurdles are formidable on the hardware side, there are strong efforts underway to be ready with suitable algorithms for nuclear applications.
These applications include calculating nuclear ground-state energies and nuclear dynamics within linear response theory~\cite{Roggero:2020sgd}, and addressing entanglement in quantum field theory, which could be relevant for both effective field theory and lattice QCD~\cite{Cloet:2019wre,Robin:2020aeh}.

\textbf{Open questions.}
The list of exploratory efforts based on emerging technologies is impressive but also invites the question: are there additional opportunities that our community has not yet explored? 
And how do we keep aware of new developments?
On the other hand, we also need to ask how much manpower and resources should be invested in areas such as quantum computing, for which noisy computers will be a fact of life for at least the short term.
These are questions best addressed by community dialog, such as facilitated by the INT Crossroads program.

\section{Coda: intersections with other fields}

The legacy from Weinberg's seminal papers on nuclear EFT has spread beyond the immediate goal of understanding nuclei.
Nuclear structure is also at crossroads with other subfields of physics in the sense of meeting places where ideas, methods, and results are shared.
These intersections include multi-messenger astronomy where gravitational wave and electromagnetic observations are confronting predictions of the neutron matter equation of state from \chiEFT; beyond-the-standard-model physics including neutrino experiments, precision tests of fundamental symmetries, and dark-matter direct detection, where EFT plays a key role in the planning and analysis of experiments; and ultra-cold atom experiments, which continue to be a testing ground for universal few- and many-body EFT calculations.
Carrying out the plans and resolving the challenges identified at the ``Nuclear Structure at the Crossroads'' program and summarized here will be crucial for the future success of this cross-disciplinary research.

\section{Workshop participants}
\label{sec:participants}

The participants in the INT program ``Nuclear Structure at the Crossroads'' were:
Bijaya Acharya, 
Anatoli Afanasjev, 
Sinya Aoki, 
Sonia Bacca, 
Jaber Balalhabashi,
Aaina Bansal, 
Carlo Barbieri, 
Nir Barnea, 
Bruce Barrett, 
Paulo Bedaque, 
George Bertsch, 
Scott Bogner, 
Antoine Boulet, 
To{\~n}o Coello P\'erez, 
Lorenzo Contessi, 
Will Detmold, 
Jacek Dobaczewski, 
Christian Drischler, 
Thomas Duguet, 
Andreas Ekstr{\"o}m, 
Jon Engel,
Dick Furnstahl, 
Hana Gil, 
Hans-Werner Hammer, 
Sophia Han, 
Wick Haxton, 
Kai Hebeler, 
Heiko Hergert, 
Emiko Hiyama, 
Martin Hoferichter, 
Jason Holt, 
Chuck Horowitz, 
Marc Illa, 
Weiguang Jiang, 
Alejandro Kievsky, 
Sebastian K{\"o}nig, 
Hermann Krebs, 
Denis Lacroix, 
Diego Lonardoni, 
Bingwei Long, 
Alessandro Lovato, 
Larry McLerran, 
Javier Men\'endez, 
Amy Nicholson, 
Thomas Papenbrock, 
Assumpta Parre{\~n}o, 
Saori Pastore, 
Jorge Piekarewicz, 
Lucas Platter, 
Caroline Robin, 
Robert Roth, 
Gautam Rupak, 
Mario S{\'a}nchez S{\'a}nchez, 
Christiane Schmickler, 
Achim Schwenk, 
Phiala Shanahan, 
Ragnar Stroberg, 
Ingo Tews, 
Bira van Kolck, 
Michael Wagman, 
Ronen Weiss, 
Kyle Wendt,
Corbinian Wellenhofer, 
Sarah Wesolowski, 
Roland Wirth, 
Jiangming Yao, 
and
Xilin Zhang.
We are grateful for their active and enthusiastic contributions to the discussions summarized here.

\begin{acknowledgements}

The work of RJF was supported by the National Science Foundation under Grant No.~PHY--1913069, and the NUCLEI SciDAC Collaboration under US Department of Energy MSU subcontract RC107839-OSU\@. The work of H.W.H. and A.S. was supported by the Deutsche Forschungsgemeinschaft (DFG, German Research Foundation) -- Project-ID 279384907 -- SFB 1245 and by the BMBF Contract No.~05P18RDFN1. We thank the Institute for Nuclear Theory at the University of Washington for its kind hospitality and stimulating research environment. The INT program ``Nuclear Structure at the Crossroads'' was supported by the INT's U.S.~Department of Energy grant No.~DE-FG02-00ER41132 and by the SFB 1245.

\end{acknowledgements}

\bibliographystyle{spphys}
\bibliography{crossroads.bib}
\end{document}